\documentclass[twocolumn]{aastex631}

\usepackage{indentfirst}
\usepackage{appendix}
\usepackage{makecell}
\usepackage{tabu}
\usepackage{float}

\usepackage{xcolor}
\usepackage{enumitem}
\usepackage{amsmath}
\usepackage{xspace}

\newcommand{\Msun}{\ensuremath{\mathrm{M}_\odot}\xspace}
\newcommand{\MBH}{\ensuremath{M_\mathrm{BH}}\xspace}
\newcommand{\MS}{\ensuremath{M_*}\xspace}

\shortauthors{Silverman et al.}

\begin{document}


\title{SHELLQs-JWST perspective on the intrinsic mass relation between supermassive black holes and their host galaxies at $z>6$}


\author{John D. Silverman}
\email{silverman@ipmu.jp}
\affiliation{Kavli Institute for the Physics and Mathematics of the Universe (Kavli IPMU, WPI), UTIAS, Tokyo Institutes for Advanced Study, University of Tokyo, Chiba, 277-8583, Japan}
\affiliation{Department of Astronomy, Graduate School of Science, The University of Tokyo, 7-3-1 Hongo, Bunkyo, Tokyo 113-0033, Japan}
\affiliation{Center for Data-Driven Discovery, Kavli IPMU (WPI), UTIAS, The University of Tokyo, Kashiwa, Chiba 277-8583, Japan}
\affiliation{Center for Astrophysical Sciences, Department of Physics \& Astronomy, Johns Hopkins University, Baltimore, MD 21218, USA}

\author{Junyao Li}
\affiliation{Department of Astronomy, University of Illinois at Urbana-Champaign, Urbana, IL 61801, USA}

\author{Xuheng Ding}
\affiliation{School of Physics and Technology, Wuhan University, Wuhan 430072, China}

\author[0000-0003-2984-6803]{Masafusa Onoue}
\affiliation{Waseda Institute for Advanced Study (WIAS), Waseda University, 1-21-1, Nishi-Waseda, Shinjuku, Tokyo 169-0051, Japan}
\affiliation{Kavli Institute for the Physics and Mathematics of the Universe (Kavli IPMU, WPI), UTIAS, Tokyo Institutes for Advanced Study, University of Tokyo, Chiba, 277-8583, Japan}

\author{Michael A. Strauss}
\affiliation{Department of Astrophysical Sciences, Princeton University, 4 Ivy Lane, Princeton, NJ 08544, USA}

\author{Yoshiki Matsuoka}
\affiliation{Research Center for Space and Cosmic Evolution, Ehime University, 2-5 Bunkyo-cho, Matsuyama, Ehime 790-8577, Japan}

\author{Takuma Izumi}
\affiliation{National Astronomical Observatory of Japan, 2-21-1 Osawa, Mitaka, Tokyo 181-8588, Japan}
\affiliation{Department of Astronomy, Graduate School of Science, The University of Tokyo, 7-3-1 Hongo, Bunkyo, Tokyo 113-0033, Japan}
\affiliation{Graduate Institute for Advanced Studies, SOKENDAI, 2-21-1 Osawa, Mitaka, Tokyo 181-8588, Japan}
\affiliation{Kagoshima University, Kagoshima 890-0065, Japan}

\author{Knud Jahnke}
\affiliation{Max Planck Institute for Astronomy, K\"onigstuhl 17, D-69117 Heidelberg, Germany}

\author{Tommaso Treu}
\affiliation{Department of Physics and Astronomy, University of California Los Angeles, CA, 90095, USA}

\author{Marta Volonteri}
\affiliation{Institut d'Astrophysique de Paris, Sorbonne Universit\'e, CNRS, UMR 7095, 98 bis bd Arago, 75014 Paris, France}

\author[0000-0002-2099-0254]{Camryn L. Phillips}
\affiliation{Department of Astrophysical Sciences, Princeton University, 4 Ivy Lane, Princeton, NJ 08544, USA}

\author[0000-0001-6102-9526]{Irham T. Andika}
\affiliation{Technical University of Munich, TUM School of Natural Sciences, Department of Physics, James-Franck-Str. 1, D-85748 Garching, Germany}
\affiliation{Max-Planck-Institut f\"{u}r Astrophysik, Karl-Schwarzschild-Str. 1, D-85748 Garching, Germany}

\author[0000-0003-4569-1098]{Kentaro Aoki}
\affiliation{Subaru Telescope, National Astronomical Observatory of Japan, 650 North A'ohoku Place, Hillo HI 96720 USA}

\author[0009-0007-0864-7094]{Junya Arita}
\affiliation{Department of Astronomy, Graduate School of Science, The University of Tokyo, 7-3-1 Hongo, Bunkyo, Tokyo 113-0033, Japan}

\author[0000-0002-9850-6290]{Shunsuke Baba}
\affiliation{Institute of Space and Astronautical Science (ISAS), Japan Aerospace Exploration Agency (JAXA), 3-1-1 Yoshinodai, Chuo-ku, Sagamihara, Kanagawa 252-5210, Japan}

\author[0000-0001-8582-7012]{Sarah E. I. Bosman}
\affiliation{Institute for Theoretical Physics, Heidelberg University, Philosophenweg 12, D–69120, Heidelberg, Germany}
\affiliation{Max-Planck-Institut f\"{u}r Astronomie, K\"{o}nigstuhl 17, 69117 Heidelberg, Germany}

\author[0000-0003-2895-6218]{Anna-Christina Eilers}
\affiliation{MIT Kavli Institute for Astrophysics and Space Research, Massachusetts Institute of Technology, Cambridge, MA 02139, USA}

\author{Xiaohui Fan}
\affiliation{Steward Observatory, University of Arizona, 933 N Cherry Avenue, Tucson, AZ 85721, USA}

\author[0000-0001-7201-5066]{Seiji Fujimoto}
\affiliation{Department of Astronomy, The University of Texas at Austin, Austin, TX 78712, USA}
\affiliation{David A. Dunlap Department of Astronomy and Astrophysics, University of Toronto, 50 St. George Street, Toronto, Ontario, M5S 3H4, Canada}

\author{Melanie Habouzit}
\affiliation{Department of Astronomy, University of Geneva, Chemin d'Ecogia, CH-1290 Versoix, Switzerland}

\author[0000-0003-3633-5403]{Zoltan Haiman}
\affiliation{Institute of Science and Technology Austria (ISTA), Am Campus 1, 3400 Klosterneuburg, Austria}
\affiliation{Department of Astronomy, Columbia University, New York, NY 10027, USA}
\affiliation{Department of Physics, Columbia University, New York, NY 10027, USA}

\author[0000-0001-6186-8792]{Masatoshi Imanishi}
\affiliation{National Astronomical Observatory of Japan, 2-21-1 Osawa, Mitaka, Tokyo 181-8588, Japan}

\author[0000-0001-9840-4959]{Kohei Inayoshi}
\affiliation{Kavli Institute for Astronomy and Astrophysics, Peking University, Beijing 100871, China}

\author[0000-0002-4923-3281]{Kazushi Iwasawa}
\affiliation{Institut de Ci\`encies del Cosmos (ICCUB), Universitat de Barcelona (IEEC-UB), Mart\'i i Franqu\`es, 1, 08028 Barcelona, Spain}
\affiliation{ICREA, Pg Llu\'is Companys 23, 08010 Barcelona, Spain}

\author[0000-0003-3954-4219]{Nobunari Kashikawa}
\affiliation{Department of Astronomy, Graduate School of Science, The University of Tokyo, 7-3-1 Hongo, Bunkyo, Tokyo 113-0033, Japan}
\affiliation{Research Center for the Early Universe, Graduate School of Science, The University of Tokyo, 7-3-1 Hongo, Bunkyo-ku, Tokyo 113-0033, Japan}

\author[0000-0002-3866-9645]{Toshihiro Kawaguchi}
\affiliation{Graduate School of Science and Engineering, University of Toyama, Gofuku 3190, Toyama 930-8555, Japan}

\author[0000-0003-1700-5740]{Chien-Hsiu Lee}
\affiliation{Hobby-Eberly Telescope, McDonald Observatory, UT Austin, 32 Mount Fowlkes, Fort Davis, TX 79734, USA}

\author[0000-0001-6106-7821]{Alessandro Lupi}
\affiliation{Como Lake Center for Astrophysics,  DiSAT, Università degli Studi dell’Insubria, via Valleggio 11, I-22100, Como, Italy}

\author[0000-0002-7402-5441]{Tohru Nagao}
\affiliation{Research Center for Space and Cosmic Evolution, Ehime University, 2-5 Bunkyo-cho, Matsuyama, Ehime 790-8577, Japan}
\affiliation{Amanogawa Galaxy Astronomy Research Center, Kagoshima University, 1-21-35 Korimoto, Kagoshima 890-0065, Japan}

\author[0000-0002-4544-8242]{Jan-Torge Schindler}
\affiliation{Hamburger Sternwarte, University of Hamburg, Gojenbergsweg 112, D-21029 Hamburg, Germany}

\author[0000-0001-7825-0075]{Malte Schramm}
\affiliation{Universit\"{a}t Potsdam, Karl-Liebknecht-Str. 24/25, D-14476 Potsdam, Germany}

\author[0000-0002-2597-2231]{Kazuhiro Shimasaku}
\affiliation{Department of Astronomy, Graduate School of Science, The University of Tokyo, 7-3-1 Hongo, Bunkyo, Tokyo 113-0033, Japan}
\affiliation{Research Center for the Early Universe, Graduate School of Science, The University of Tokyo, 7-3-1 Hongo, Bunkyo-ku, Tokyo 113-0033, Japan}

\author[0000-0002-3531-7863]{Yoshiki Toba}
\affiliation{Department of Physical Sciences, Ritsumeikan University, Kusatsu, Shiga 525-8577, Japan}
\affiliation{Academia Sinica Institute of Astronomy and Astrophysics, 11F of Astronomy-Mathematics Building, AS/NTU, No.1, Section 4, Roosevelt Road, Taipei 10617, Taiwan}
\affiliation{Research Center for Space and Cosmic Evolution, Ehime University, 2-5 Bunkyo-cho, Matsuyama, Ehime 790-8577, Japan}

\author[0000-0002-3683-7297]{Benny Trakhtenbrot}
\affiliation{School of Physics and Astronomy, Tel Aviv University, Tel Aviv 69978, Israel}
\affiliation{Max-Planck-Institut f{\"u}r extraterrestrische Physik, Gie\ss{}enbachstra\ss{}e 1, 85748 Garching, Germany}
\affiliation{Excellence Cluster ORIGINS, Boltzmannsstra\ss{}e 2, 85748, Garching, Germany}

\author{Hideki Umehata}
\affiliation{Institute for Advanced Research, Nagoya University, Furocho, Chikusa, Nagoya 464-8602, Japan}
\affiliation{Department of Physics, Graduate School of Science, Nagoya University, Furocho, Chikusa, Nagoya 464-8602, Japan}

\author[0000-0001-9191-9837]{Marianne Vestergaard}
\affiliation{DARK, The Niels Bohr Institute, University of Copenhagen, Jagtvej 155, 2200 Copenhagen N, Denmark}
\affiliation{Steward Observatory, University of Arizona, 933 N Cherry Avenue, Tucson, AZ 85721, USA}

\author{Fabian Walter}
\affiliation{Max-Planck-Institut f\"{u}r Astronomie, K\"{o}nigstuhl 17, D-69117 Heidelberg, Germany}

\author[0000-0002-7633-431X]{Feige Wang}
\affiliation{Department of Astronomy, University of Michigan, 1085 S. University Ave., Ann Arbor, MI 48109, USA}

\author[0000-0001-5287-4242]{Jinyi Yang}
\affiliation{Department of Astronomy, University of Michigan, 1085 S. University Ave., Ann Arbor, MI 48109, USA}



\begin{abstract}

The relation between the masses of supermassive black holes (SMBHs) and their host galaxies encodes information on their mode of growth, especially at the earliest epochs. The James Webb Space Telescope (JWST) has opened such investigations by detecting the host galaxies of AGN and more luminous quasars within the first billion years of the universe ($z\gtrsim6$). Here, we evaluate the relation between the mass of SMBHs and the total stellar mass of their host galaxies using a sample of nine quasars at $6.18\leq z \leq 6.4$ from the Subaru High-$z$ Exploration of Low-luminosity Quasars (SHELLQs) survey with NIRCam and NIRSpec observations.  We find that the observed location of these quasars in the SMBH--galaxy mass plane ($\log M_\mathrm{BH}/\Msun \sim8$--9; $\log M_*/\Msun \sim9.5$--11)  is consistent with a non-evolving intrinsic mass relation with dispersion ($0.80_{-0.28}^{+0.23}$\,dex) higher than the local value ($\sim$\,0.3--0.4\,dex) of their more massive descendants. Our analysis is based on a forward model of systematics and includes a consideration of the impact of selection effects and measurement uncertainties with an assumption on the slope of the mass relation. While degeneracies between parameters persist, the best-fit solution has a reasonable AGN fraction (2.3\%) of galaxies at $z\sim6$ with an actively growing UV-unobscured black hole. In particular, models with a substantially higher normalisation in \MBH would require an unrealistically low intrinsic dispersion ($\sim$\,0.22\,dex). Consequently, our results predict a large population of AGNs at lower black hole masses, as are now just starting to be discovered in focused efforts with JWST.   

\end{abstract}



\keywords{}






\section{Introduction}



The James Webb Space Telescope (JWST; \citealt{Rigby2023}) has opened the landscape of the early universe and the formation of the first SMBHs \citep{Inayoshi2020,Volonteri2021}. Deep JWST survey fields routinely find SMBHs with a mass estimated to be $\MBH\sim10^{7\text{--}8}$\,\Msun\citep[e.g.,][]{Onoue2023,Lupi2024} and identified as Little Red Dots (LRDs) in many cases \citep[e.g.,][]{Matthee2024,greene2024,Kocevski2024,Maiolino2024b}. In particular, an even lower mass ($\MBH\sim10^{6}$\,M$_{\odot}$) black hole may lurk in GN-z11 at $z=10.6$ \citep{Maiolino2024a}. It is imperative to understand their growth history and relation to the more massive black holes at $z\sim6$ which are observed as luminous quasars.



Astronomers have been making remarkable breakthroughs with JWST in the study of quasars and their massive host galaxies at high redshift. Starlight from quasar host galaxies is now often detected at $z>6$ \citep{Ding2023,Stone2023,Stone2024,Yue2024,Onoue2024,LiHo2025}. This is a giant leap from the pre-JWST era when one could study the stellar properties of hosts routinely only to $z\sim2$ \citep{Jahnke2004,Mechtley2016,Ding2022b}, or in a handful of lensed cases \citep{Peng2006} and mergers \citep{Decarli2019} at higher redshifts.  

The Subaru High-$z$ Exploration of Low-luminosity Quasars (SHELLQs) sample \citep[e.g.,][]{Matsuoka2022}, constructed from the wide and deep imaging survey with Subaru's Hyper Suprime-Cam \citep{Aihara2018,Miyazaki2018}, is playing an important role in bridging the gap between luminous quasars ($M_\mathrm{UV}<-25$) and the fainter JWST-discovered AGNs ($M_\mathrm{UV}>-22$) at $z\sim6$. Based on JWST followup with NIRCam, the hosts of SHELLQs quasars have high stellar light fractions (up to 70\% in F356W; \citealt{Ding2023, Ding2025}) that enable robust measurements of galaxy sizes (0.6--3.0\,kpc) and stellar masses ($\MS>10^{10}\,\Msun$) with typical precision of 0.3\,dex. With NIRSpec, black hole masses are determined to be between $10^{7.8}$ and $10^{9.1}$\,\Msun based on the broad ($\mathrm{FWHM} > 2000$\,km\,s$^{-1}$) Balmer emission lines (\citealt{Ding2023, Matsuoka2025}, Onoue et al.\ in preparation). Furthermore, these observations reveal that some quasar hosts exhibit post-starburst signatures (i.e., Balmer absorption lines, Balmer breaks, and lack of spatially extended H$\alpha$ emission; \citealt{Onoue2024}) suggesting that quasar activity may be tied with quenching in the early universe. 

The relation between black hole mass, $\MBH$, and host stellar mass, \MS, over cosmic time depends on the connections between the physics of galaxy and black hole growth. Hydrodynamic simulations of galaxy formation vary in their prediction of the relative mass growth of  SMBHs and their host galaxies, likely due to differences in black hole seeding, modeling of black hole accretion, and feedback prescriptions for stars, supernova and AGN \citep{Habouzit2022}. Observationally, there appears to be little deviation from the local \MBH--\MS relation up to $z\sim2.5$ \citep[e.g., ][]{Cisternas2011,Sun2015,Shen2015,Ding2020,Li2021mass,Tanaka2024}. However, our current best estimates on the evolution of the mass relation at high redshift depends on accurate accounting of selection biases, measurement uncertainties and our relatively limited knowledge of the Eddington rate distribution at high redshifts. Even if systematics are under control, there remains a degeneracy between the evolution rate and the intrinsic scatter, even with the large samples compiled from wide-field ground-based studies up to $z\sim1$ \citep{Li2021mass}. Pushing observational studies to higher redshift ($z\sim6$), closer to their formation epoch, offers much promise since hydrodynamic simulations show larger deviations \citep{Habouzit2022}, even though similar biases and selection effects are still present \citep[e.g.,][]{Schulze2014,Li2025b} and seen in ALMA studies based on dynamical masses for their hosts \citep[e.g.,][]{Izumi2019,Li2022,Wang2024}.    

The initial results from JWST on AGN and quasars at $z\gtrsim4$ provide new insights into the mass relation. To date the majority of observed AGNs and quasars at high-$z$ \citep[e.g.,][]{Maiolino2024b,Yue2024,Yang2025,Marshall2025,Jones2025}, with measured stellar masses, have black hole masses higher than predicted from the local mass relation \citep{Kormendy2013,Reines2015,Greene2020}. This is anticipated when considering the effects of selection biases mentioned above. \citet{Ding2023} find that the first two quasars with detected hosts at $z>6$ lie in a region of the \MBH--\MS plane as expected when assuming the local relation and proper treatment of selection effects. For lower luminosity/mass black holes found by JWST at high-$z$, 
\MBH/\MS has been claimed to lie above the local relation \citep{Pacucci2023,Maiolino2024b}. This has been interpreted as a sign for a `heavy seed' formation channel, likely required to explain many of the individual observed cases. However, it has been demonstrated by \citet{Li2025b} that selection effects are important even at lower masses when inferring the intrinsic relation, which requires detailed modeling of the expected underlying population of SMBHs and their host galaxies, in the absence of detected systems on or below the local relation. Recently, \citet{Li2025a} have demonstrated that such a predicted population of AGN with lower mass black holes is found in massive galaxies with JWST spectroscopy. \citet{Geris2025} further identified a population of low-mass black holes in low-mass galaxies at $3<z<7$ located within the scatter of the local relation by stacking deep NIRSpec spectroscopy. Similar cases are being revealed in star-forming main-sequence galaxies at $4<z<6$ from the ALPINE/CRISTAL surveys using JWST NIRSpec IFU (Ren et al., in preparation; Faisst et al. in preparation).

Here, we report on the relation between the mass of SMBHs and their host galaxies at $z>6$ using a Cycle~1 JWST program which observed twelve low-luminosity quasars from the SHELLQs program (GO 1967, PI: M. Onoue). NIRCam imaging provides the host galaxy detection after careful modeling and subtraction of the quasar emission, while spectroscopy with NIRSpec enables a measure of the black hole mass from the velocity-broadened Balmer emission lines. The details of each of these efforts will be fully reported separately (\citealt{Ding2025}; Onoue et al. in preparation). We use the local black hole to host galaxy mass relation of all galaxy types from \citet{Greene2020} based on dynamically measured black hole masses as a basis for comparison with the high-redshift results. This relation is treated as representative of the overall black hole population in the local Universe. The local relations established for AGNs \cite[e.g.,][]{Reines2015} are likely biased toward lower-mass black holes due to an additional AGN duty cycle-dependent selection bias \citep{Schulze2011, Li2025a}, and are therefore not used in our comparison. Throughout this work, we use a Hubble constant of $H_0 = 70$ km s$^{-1}$ Mpc$^{-1}$ and cosmological density parameters $\Omega_\mathrm{m} = 0.3$ and $\Omega_\Lambda = 0.7$. We assume a \citet{Chabrier2003} stellar initial mass function to infer the host galaxy masses.

\begin{figure}
\epsscale{1.2}
\plotone{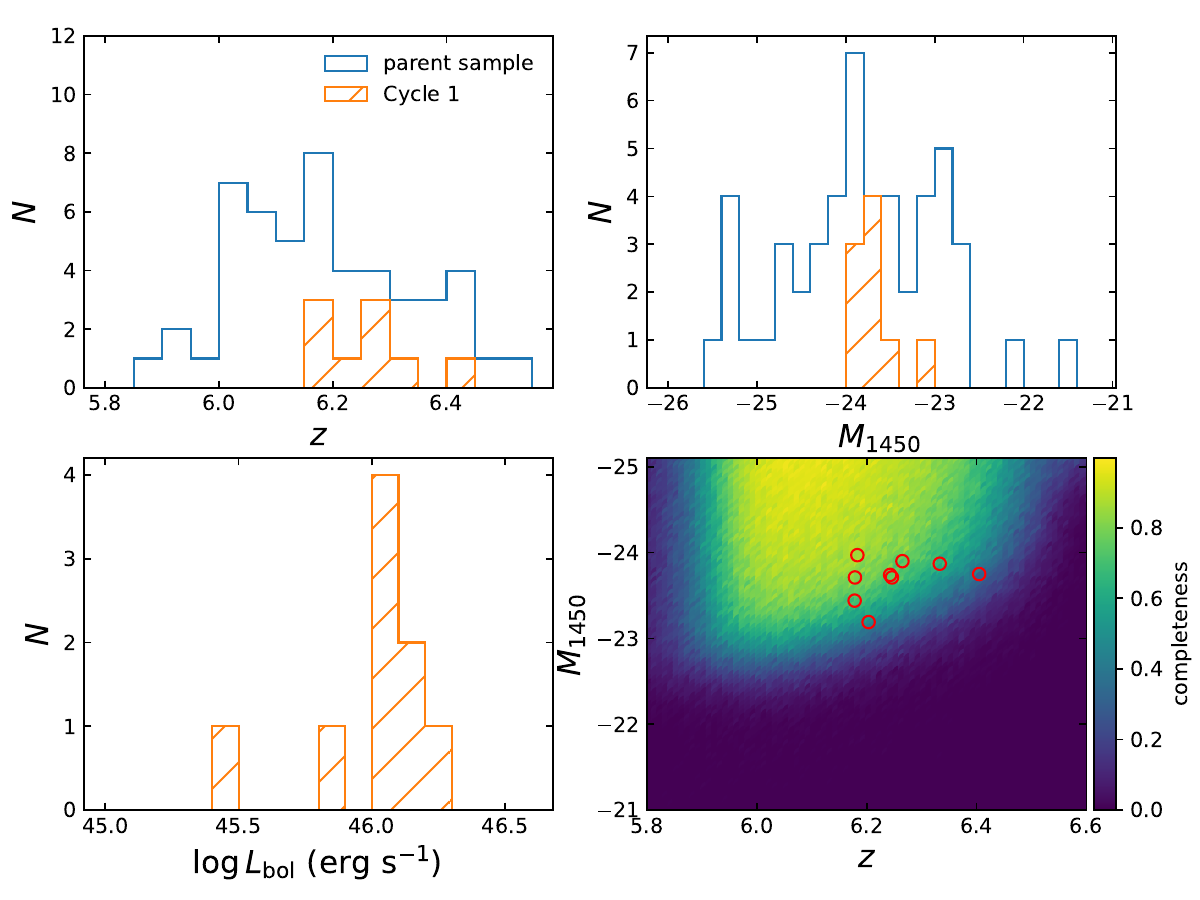}
\caption{Properties of the Cycle 1 sample as drawn from SHELLQs \citep{Matsuoka2018ApJ}: redshift, absolute UV magnitude, bolometric luminosity, and selection completeness as a function of absolute magnitude and redshift.}
\label{fig:sample}
\end{figure}

\section{Data and measurements}

\subsection{Sample selection}

Twelve quasars were selected from an early sample of 50 low-to-moderate luminosity ($M_{1450}>-26$ mag) SHELLQs quasars discovered by Subaru's Hyper Suprime-Cam \citep{Aihara2018} over 650\,deg$^2$, spectroscopically confirmed from ground-based observations (mainly by their Ly$\alpha$ emission; \citealt{Matsuoka2016,Matsuoka2018PASJ,Matsuoka2018ApJS}), and included in the luminosity function reported in \citet{Matsuoka2018ApJ}. The properties of these quasars are listed in Table 1 of \citet{Ding2025} and shown in Figure~\ref{fig:sample} along with the selection completeness, ranging from 42 to 86\%. Their redshifts are between 6.18 and 6.40, on the high end of the parent SHELLQs sample \citep{Matsuoka2022}. These quasars were chosen to have faint absolute magnitudes ($-24<M_{1450}<-21$) to minimize the challenge of detecting the host galaxy under the emission from an unresolved bright point source. For this study, we remove the three faintest quasars with $M_{1450}
\gtrsim -22$  (HSC J0911+0152, HSC J1146-0005 and HSC J1512+4422) since they were added to the JWST program with a different selection from the brighter targets, i.e., not from the sample of \citet{Matsuoka2018ApJ}. 
The final sample has 9 quasars with $-24<M_{1450}<-23$. For each, we have made an assessment of the completeness (bottom-right panel of Fig.~\ref{fig:sample}), with respect to being selected from HSC photometry as a high-$z$ quasar by a Bayesian criteria, using a large mock sample with SDSS-like quasar spectra, a modified Lyman-$\alpha$ equivalent-width distribution to match HSC quasars, and IGM absorption as fully detailed in Section 3.4 of \citet{Matsuoka2018ApJ}. 

\subsection{Stellar and black hole masses with JWST}

NIRCam and NIRSpec observations of these nine SHELLQs quasars were taken in Cycle 1 as fully described by \citet{Ding2023,Ding2025} and \cite{Onoue2024}. The primary aims were to measure the stellar mass of the host galaxy using NIRCam and the mass of the black hole responsible for the quasar emission with NIRSpec.

NIRCam images of each quasar are taken in the F356W ($\lambda_\mathrm{rest}\sim0.488\,\micron$) and F150W ($\lambda_\mathrm{rest}\sim0.206\,\micron$) filters. The pixel scales are 0\farcs0315 (F356W) and 0\farcs0153 (F150W). For each image, the emission is fit, using Galight \citep{Ding2021}, with a two-component model, representing the quasar and its host galaxy. The quasar component is based on a characterization of the PSF \citep{Ding2025} while the host galaxy is represented as a S\'{e}rsic function. Briefly, the host galaxies are detected in the F356W filter for all nine quasars based on satisfying three conditions: a host-to-total flux ratio greater than 3\% (based on extensive simulations of quasar+host decomposition), signal-to-noise of the host ($\ge2$), and improvement in the Bayesian Information Criterion ($\Delta \mathrm{BIC} > 50$) when including the host component in the modeling \citep{Ding2025}. The host-to-total ratios vary from $\sim4\%$ to $\sim69\%$. Six of the nine quasars are also detected in the F150W filter and each are significantly fainter with lower signal-to-noise ratio. Uncertainties on best-fit parameters are assessed by varying the method and stars used to construct the PSF model.

We infer the host stellar mass by fitting the available photometry with SED models through the software package GSF \citep{Morishita2019}, which results in the range from $10^{9.5\text{--}11}$\,\Msun. Eight of the nine quasars hosts have photometry in two filters (F356W and F150W), which bracket the Balmer break at their redshifts. For J2236+0032, the host is observed with eight filters, thus providing an accurate assessment of the underlying stellar population \citep{Onoue2024} that lends weight to the robustness of results for those with only two filters. Upper limits are used for those with non-detections in F150W. In addition to a chosen \cite{Chabrier2003} IMF, the SEDs are mainly dependent on stellar age, metallicity (Z) and dust attenuation (A$_V$). Typical stellar mass errors are $0.35$\,dex, which encompass uncertainties in these parameters (age, Z and A$_V$) and are each reflective of errors in the host photometry, resulting from varying the method of 2D decomposition as mentioned above. The three quasars with single-band detections in F356W are used here with larger uncertainties ($\sim 0.4-0.5$ dex) on their stellar masses as reported in \citet{Ding2025}, which provides a full description of the assumptions on parameters and their uncertainties.

\begin{figure}
\epsscale{2.4}
\plottwo{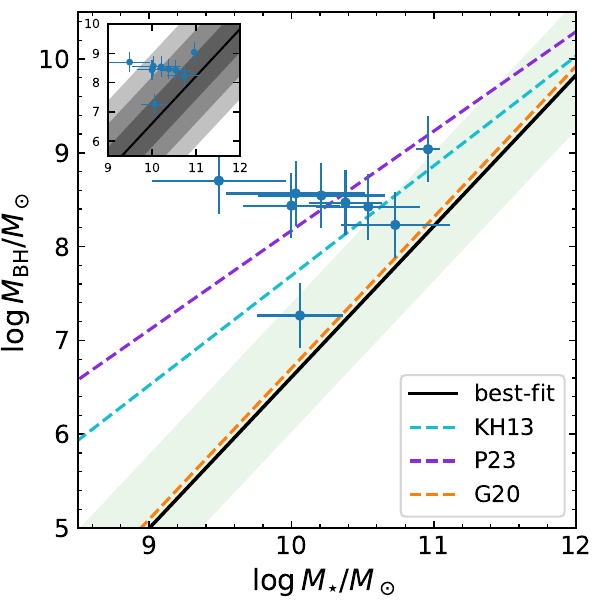}{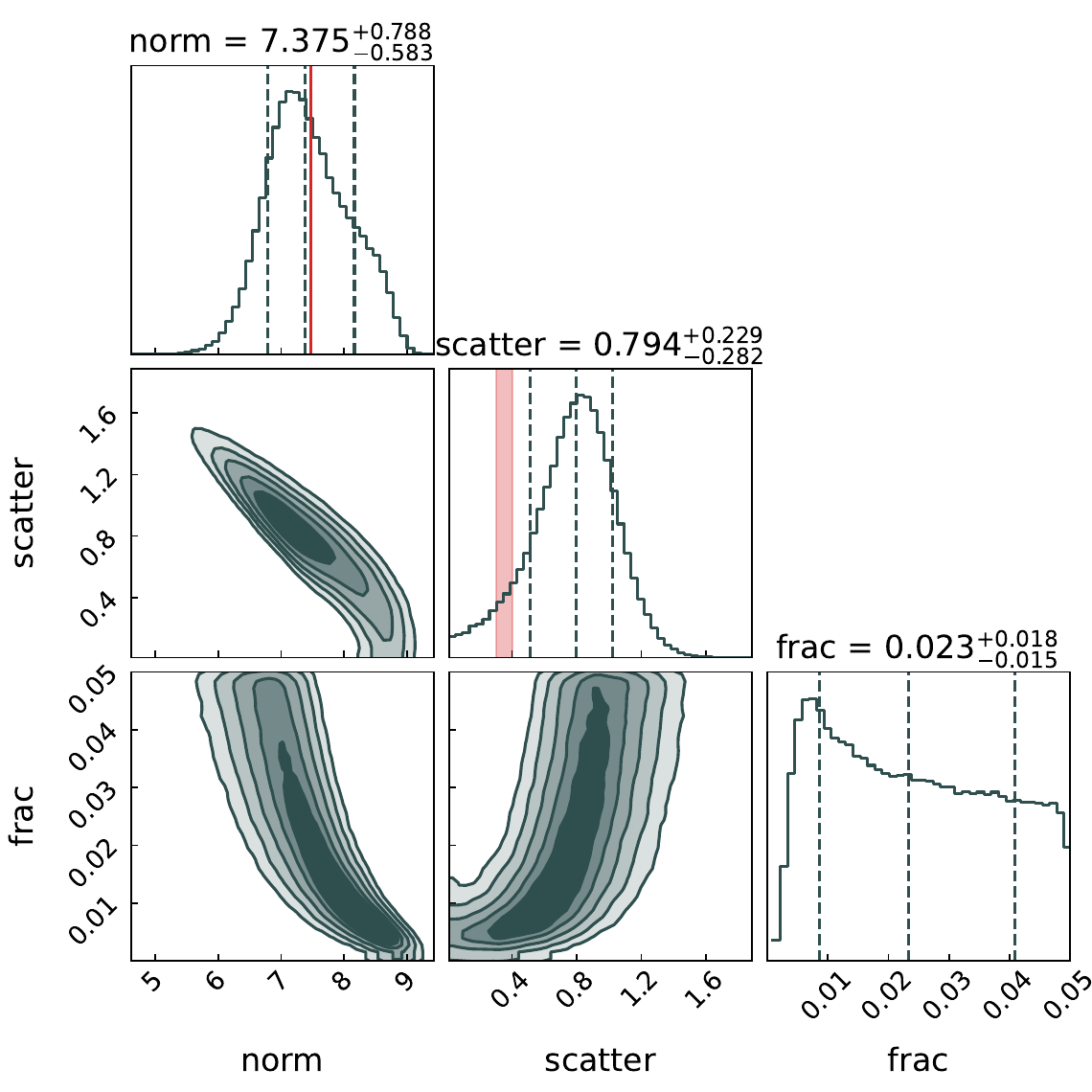}
\caption{$Top$ Black hole mass (\MBH) versus stellar mass (\MS) of the host galaxy for the 9 SHELLQs quasars. The best-fit relation (black line) and uncertainty (1$\sigma$; shaded area) are indicated which incorporate selection biases and measurement uncertainties. Local relations of \citealt{Greene2020} (G20) and \citealt{Kormendy2013} (KH13) are also indicated, along with the high-$z$ assessment of \citealt{Pacucci2023} (P23) using JWST AGN in CEERS \citep{Kocevski2024} and JADES \citep{Maiolino2024b}. The inset plot displays the location of the observed quasar sample with respect to our best-fit relation with shaded areas marking the intrinsic scatter (1--3$\sigma$). $Bottom$ Best-fit inference on model parameters (normalization, scatter of the linear mass relation, and AGN fraction) with the slope fixed to 1.61. The normalization of the G20 relation is indicated by the vertical red line in the top panel, while the local intrinsic dispersion of massive local galaxies \citep{Gueltekin2009,Kormendy2013,Greene2020,Bennert2021} is shown by the red vertical band in the top-middle panel.}
\label{fig:mass_rel}
\end{figure}

NIRSpec observations with a fixed slit and G395M grating ($R\approx1000$; $2.87\,\micron <\lambda_\mathrm{obs}< 5.27\,\micron$) provide a wealth of spectral information for understanding the physical nature of quasar emission and, if detected, the underlying host galaxy. We model the continuum and emission lines to measure the velocity widths of the broad Balmer emission lines, H$\alpha$ and H$\beta$. Here, we use the single-epoch virial method to determine black hole masses based on H$\beta$ \citep{Vestergaard2006}, the same method which is used for local scaling relations. The resulting black hole masses are found to range between $\MBH=10^{8\text{--}9}$\,\Msun, for all objects in the sample\footnote{For one quasar (J0217–0208), \citet{Marques-Chaves2025} claim that the broad lines are due to an outflow rather than emission from the AGN broad-line region. Even so, we include this object in our quasar sample.}, except one at $\MBH=10^{7.2}$\,\Msun, as shown in Figure~\ref{fig:mass_rel}. The uncertainties are at the level of 0.4\,dex \citep{Shen2024} which is taken into consideration in our analysis described below. 

\section{Method to account for selection biases and measurement uncertainties}
\label{sec:methods}
There are numerous works \citep[e.g.,][]{Treu2007,Lauer2007,Schulze2011,Schulze2014} highlighting the potential for strong biases in SMBH--host galaxy mass relations when using AGNs in flux-limited samples, particularly those extending to high redshifts. Here, we follow the recent procedure described by \citet{Li2025b}  to infer the underlying \MBH--\MS relation from the observed data points. Recently, studies have begun to apply such corrections for selection effects when evaluating the intrinsic mass relation \citep{Li2021mass,Li2022,Pacucci2023,Tanaka2024,sun2025}. 
 
We briefly describe the procedure for determining the likelihood distributions on model parameters descriptive of the intrinsic mass relation at high redshift while referring the reader to \citet{Li2025b} for the details. A forward model of the intrinsic linear \MBH--\MS relation in log space is implemented with constraints from the observed values of \MBH, \MS, and their number counts within the survey area. Our modeling requires key distribution functions describing the underlying true galaxy and black hole populations.

Explicitly, a functional form of the conditional probability distribution for detecting a quasar is constructed as a function of the observed stellar mass, black hole mass (assuming gaussian uncertainties), bolometric luminosity, emission-line width, active unobscured AGN fraction and redshift \citep[see Equation 4 in][]{Li2025b}. First, we start with an analytic form of the stellar mass function of galaxies as a function of redshift from \citet{Shuntov2025}. SMBHs are connected to galaxies through the \MBH--\MS mass relation which is described by its slope ($\beta$), normalization ($\alpha$), and intrinsic scatter ($\sigma$). We parameterize the linear mass relation following \cite{Greene2020} as:
\begin{equation}
    {\rm log}\,\MBH/M_\odot = \alpha + \beta\, {\rm log}\,(\MS/M_0) + \sigma
\end{equation}
where $M_0 = 3\times10^{10}$ M$_{\odot}$ and $\sigma$ is the intrinsic Gaussian scatter, i.e., not including measurement errors. The Eddington rate distribution function (ERDF) provides the means to assign a likelihood distribution on the bolometric quasar luminosity which then determines whether an individual quasar is luminous enough to be included in the SHELLQs sample. We implement the log-normal ERDF at $z\sim6$ provided by \citet{Wu2022} which peaks at $\log \lambda_\mathrm{Edd}\sim-1.0$ with a scatter of $\sim0.4$. The unobscured broad-line AGN fraction indicates the population of galaxies with actively accreting SMBHs following the assumed ERDF and selection as \textit{UV-unobscured} quasars by SHELLQs. To be conservative, we constrained this fraction (p$_\mathrm{active}$) to be less than $5\%$ assuming a flat prior, motivated by recent clustering-based estimates of the duty cycle of SHELLQs quasars ($\sim1.9\%$; \citealt{Arita2023}). Adopting higher values (e.g., up to $\sim20\%$) based on recent studies of JWST optically selected broad-line AGNs \citep{Harikane2023,Maiolino2024a,Li2025b} would shift the recovered intrinsic mass relation (Sec.~\ref{sec:result}) further downward in the \MBH--\MS plane to avoid overproducing detectable quasars. Nevertheless, this does not affect our conclusion that a large population of low mass black holes is likely missed (Sec.~\ref{sec:result}).

The selection function is mainly based on the quasar limiting luminosity above which a photometrically-selected quasar would be targeted for spectroscopic followup. A range in luminosity ($46 < \log L_\mathrm{bol}/(\rm {erg\,s^{-1} })< 46.2$), is imposed which is representative of the Cycle 1 sample (Figure~\ref{fig:sample}). Bolometric luminosities are computed using a correction factor from \citet{Richards06} with $L_\mathrm{bol}=9.26\ L_{5100}$. 
There is a further restriction on the broad-line widths to be greater than 2000 km s$^{-1}$. Equations 5--7 in \citet{Li2025b} give the formulation for calculating the selection function, the bivariate distribution function of observed masses with measurement uncertainties, and the derivation of the observed mass relation, expected from the modeling given the selection function.

The modeling procedure presented by \cite{Li2025b} requires a comparison of the model-predicted number counts of detectable quasars, within the observed stellar mass ranges, with the actual detected numbers (i.e., ensuring the proper normalization) to determine the best-fit model assuming Poisson likelihood. We first apply a kernel density estimate to our sparse stellar mass distribution (corrected using the incompleteness in Figure \ref{fig:sample}) to estimate the number of quasars in the following stellar mass bins: $\log\MS/\Msun < 9.5$, $9.5 < \log\MS/\Msun < 10.0$, $10.0 < \log\MS/\Msun < 10.5$, $10.5 < \log\MS/\Msun < 11.0$, and $\log\MS/\Msun > 11.0$, yielding 0.7, 2.7, 3.4, 3.9, and 1.1 sources, respectively. Since the Cycle 1 sample represents only a subset of quasars detected by SHELLQs, we further use the bolometric luminosity function of SHELLQs quasars, converted from the UV luminosity function in \cite{Matsuoka2018ApJ} assuming $L_{\rm bol} = 3.81 \times L_{\rm 1450}$ \citep{Richards06}, to estimate the total number of quasars at $\log L_{\rm bol} \sim 46.1$ (units of erg\,s$^{-1}$). This results in a correction factor of $\sim1.4$ (i.e., ratio of true quasars to those observed by JWST). In our fitting procedure, we assume that these missing objects follow the same \MBH--\MS relation as those in the Cycle 1 sample. Lastly, we refrain from merging samples from the literature with widely different selection functions which are a challenge to model simultaneously.

\begin{figure}
\epsscale{1.0}
\plotone{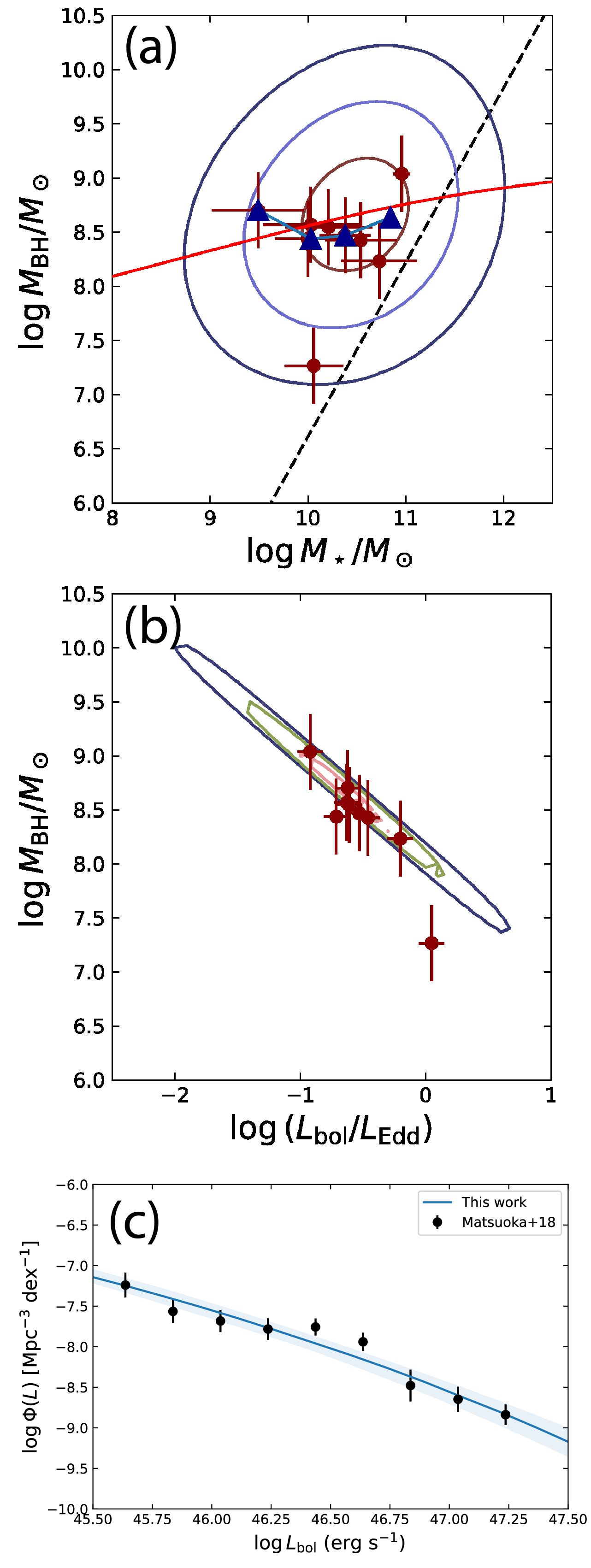}
\caption{Model comparison to the observed sample. (a) Bi-variate mass distribution ($1\sigma$ to $3\sigma$ contours). Red curve and blue triangles show the mean mass relation of the observed model distribution and the observed data for the nine SHELLQs quasars, respectively. The recovered intrinsic mass relation is shown by the dashed line. (b) Observed distribution in the \MBH vs. Eddington ratio plane with the model prediction ($1\sigma$ to $3\sigma$ contours). (c) Best-fit model luminosity function compared to that from \citet{Matsuoka2018ApJ}.}
\label{fig:LF}
\end{figure}

\section{Results}
\label{sec:result}

Top panel of Figure~\ref{fig:mass_rel} shows the location of the nine SHELLQs quasars in the \MBH--\MS plane. Their masses mostly span an interval of $8<\log \MBH / M_{\odot} <9$ and $9.5<\log \MS / M_{\odot}<11$. Similar to most quasar-selected samples at high redshift \citep[e.g.,][]{Yue2024}, the observed values fall above the local mass relations, i.e., their black hole masses are higher, given the stellar masses of their host galaxies. Their location is similar to where the two SHELLQs quasars were first reported in \cite{Ding2023} as observed by JWST. However, the observed sample is affected by selection biases and is thus not representative of the intrinsic distribution without additional inference. As demonstrated below, a proper modeling of selection effects and measurement uncertainties provide insight into the underlying intrinsic population.

Our aim here is to use the observed sample and our forward modeling procedure \citep{Li2022,Li2025b} with selection biases and measurement uncertainties incorporated to infer the underlying intrinsic mass relation between SMBHs and their hosts. As done in \citet{Li2025a}, we sample the conditional probability distributions, described in Section~\ref{sec:methods}, using \textbf{emcee} \citep{Forman-Mackey2013} to determine the most likely \MBH--\MS relation, primarily the normalization ($\alpha$) and scatter ($\sigma$) from Equation 1. Given the limited size and dynamic range in black hole mass of the SHELLQs-JWST sample, we fixed the slope ($\beta$) to 1.61 -- the value of the local relation reported in \cite{Greene2020}-- since the slope is unconstrained in our fitting and tends to hit the upper limit when left free.

In the bottom panel of Figure~\ref{fig:mass_rel}, the results are presented from the forward modeling of the \MBH--\MS relation. The likelihood distributions of each parameter are reported from left to right along the bottom axis: (1) normalization ($\alpha$) of the linear relation, (2) dispersion ($\sigma$) in the mass ratio and (3) the fraction of galaxies ($p_\mathrm{active}$) hosting UV-unobscured broad-line AGN. 

Based on the fixed slope, the best-fit intrinsic \MBH--\MS relation at $z>6$ is shown by the black line and shaded area in Fig.~\ref{fig:mass_rel} ($top$ panel) which has a normalization ($\alpha$) of $7.38_{-0.58}^{+0.79}$ at $M_{*, 0}=3\times10^{10}$\,\Msun. This is consistent with the local relation of \citet{Greene2020}, based on all galaxy types\textcolor{blue}{\footnote{We consider the \citet{Greene2020} relation for all galaxy types appropriate for comparison since many of the hosts at $z>6$ resemble disk-like galaxies \citep{Ding2025}.}}, while falling below the \MBH--$M_\mathrm{bulge}$ relation of massive bulge-dominated galaxies from \citet{Kormendy2013}. This result is inconsistent with the intrinsically overmassive relation of \citet{Pacucci2023}, shown by the dashed purple line, which is based on less massive galaxies having $7.5\lesssim \log \MS/\Msun\lesssim10$ and $6.5\lesssim \log\MBH/\Msun\lesssim8$. We note that adopting a flatter slope in the fitting, such as the value from the \cite{Kormendy2013} relation ($\beta=1.17$), does not affect our main conclusion, although it results in a significantly poorer fit.

Our model constrains the intrinsic scatter of the \MBH--\MS relation to be $0.80_{-0.28}^{+0.23}$\,dex. If we increase the uncertainties to 0.45 dex on average to our black hole mass measurements, the best-fit scatter increases to 0.85 dex. We compare the dispersion at high-$z$ with that of massive ($\log~M_*\gtrsim11$) galaxies at low-$z$, which are likely the descendants of the high-$z$ quasars. The scatter, now determined at $z>6$, is significantly higher than local massive galaxies (0.3--0.4\,dex; \citealt{Gueltekin2009,Kormendy2013,Greene2020,Bennert2021}) and low-redshift AGNs, considering spheroid \citep{Bennert2021} or total stellar \citep{Bentz2018,Reines2015} masses. We note that \citet{Greene2020} report a higher intrinsic scatter of $0.65\pm0.05$ when including lower mass galaxies down to $\sim10^{10}$ M$_{\odot}$ which warrants further investigation on the appropriate local scatter for comparison to high-$z$ estimates which is beyond the scope of this work.

As a consistency check on the model of our best-fit mass relation for an assumed ERDF, we successfully reproduce the distribution of the observed sample in both the \MBH--\MS and \MBH--Eddington ratio planes (Fig.~\ref{fig:LF}$a$--$b$). Our model constraint on the number counts of UV-unobscured quasars results in agreement with the luminosity function of \cite{Matsuoka2018ApJ} as shown in Fig.~\ref{fig:LF}$c$. This is then reflected in the best-fit unobscured active fraction ($p_\mathrm{active}=2.3_{-1.5}^{+1.8}\%$) of UV-unobscured quasars which generally agrees with the duty cycle measured through clustering analysis of quasars at $z>6$ \citep{Arita2023,Eilers2024}.

As a result, the fact that the observed SHELLQs quasars lie above the local relation in the \MBH--\MS plane is due to the dispersion in the intrinsic mass relation and the effects of selection and measurement uncertainties. While the observed quasars do have such high mass ratios individually, selection makes them biased tracers of the underlying global population that has lower mass ratios. 

\begin{figure}
\epsscale{2.4}
\plottwo{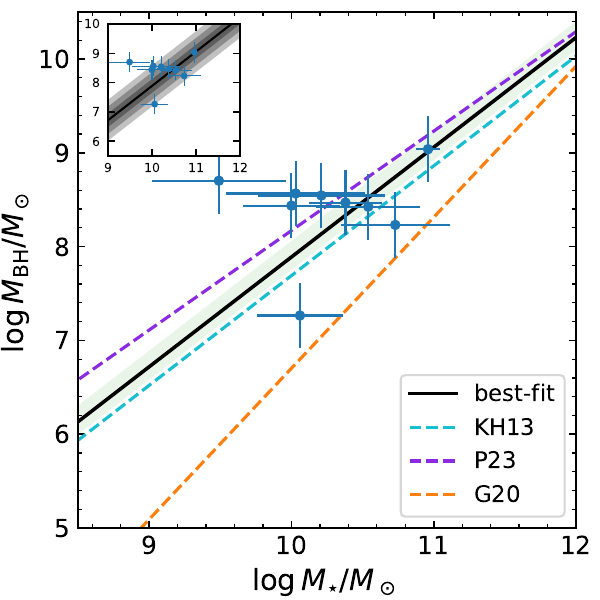}{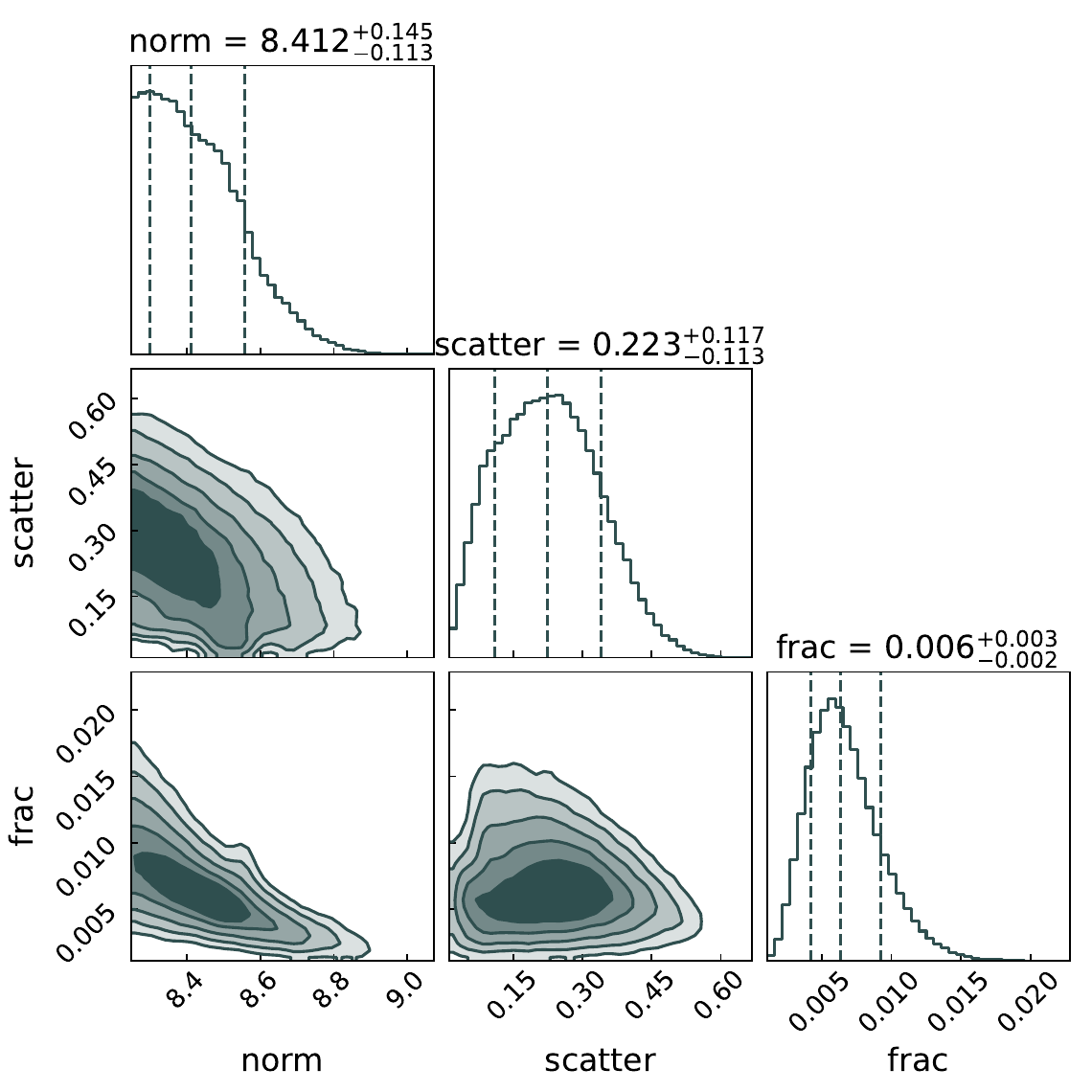}
\caption{Same as Figure~\ref{fig:mass_rel} but with the intrinsic mass relation forced to be larger than the \citet{Kormendy2013} relation (i.e., overmassive case) and the slope fixed to 1.17. As shown, this case demands a mass relation with very little dispersion (0.22, even less than the local relation) and a very low AGN fraction (0.6\%); therefore, this solution is not preferred.}
\label{fig:mass_rel_high}
\end{figure}


However, there are considerable degeneracies between parameters shown by the confidence intervals in the lower panels of Figure~\ref{fig:mass_rel}. For instance, a higher normalization results in both a lower scatter (middle left panel) and lower AGN fraction (lower left panel). Due to the limited sample size, the AGN fraction is poorly constrained within our chosen bounds (bottom right panel). While a peak is apparent in the AGN fraction at low values ($\lesssim0.01$), such a low value is unlikely since the scatter would be forced to be very low ($\sim0.2$). 

To further illustrate these effects, we show a similar set of panels in Figure~\ref{fig:mass_rel_high} for the case that requires a higher normalization (i.e., forcing a high black hole mass solution) by simply setting a prior on the lower bound of $\alpha$ to match the local value of massive ellipticals from \citet{Kormendy2013}. For this experiment, the slope ($\beta$) is fixed to that of the \citet{Kormendy2013} relation (1.17), which is close to the best-fit value when the slope range is limited to the values spanned by various local relations ($\sim$0.98-1.61; \citealt{Kormendy2013, Ding2022a, Greene2020}). The best-fit result (top panel of Fig.~\ref{fig:mass_rel_high}) is consistent with our observed quasars and closer to the best-fit relation of \citet{Pacucci2023}. However, this comes with two consequences: the scatter ($\sigma=0.22_{-0.11}^{+0.12}$ dex) and AGN fraction ($p_\mathrm{active}=0.6_{-0.2}^{+0.3}\%$) are significantly lower than the above model shown in Figure~\ref{fig:mass_rel}, in order to avoid overproducing large numbers of detectable quasars that are not observed. Intuitively, this trend can be understood as follows. With a higher normalization which falls closer to the observed data, the intrinsic scatter will be lowered since the differences between the two are effectively reduced. As mentioned above, this covariance between the scatter and normalization is clearly seen in the lower panels of Figure~\ref{fig:mass_rel}. Furthermore, this effect is amplified by the fact that some of the scatter in the data is the result of measurement uncertainties in \MBH and \MS. 

In this scenario with a higher normalization, a low intrinsic scatter of 0.22\,dex is unlikely since this value is lower than measured locally ($\sim$0.3--0.4\,dex). A higher intrinsic dispersion is expected at higher redshifts simply from cosmic averaging (i.e., central limit theorem) in merger scenarios \citep{Jahnke2011}. With the likelihood of a higher intrinsic dispersion and probably an actual AGN fraction above 1\% \citep[e.g.,][]{Arita2023}, we conclude that the intrinsic mass relation at $z\sim6$ is unlikely to be highly elevated from the local relation. 

\subsection{Comparison to simulations}

\begin{figure}
\plotone{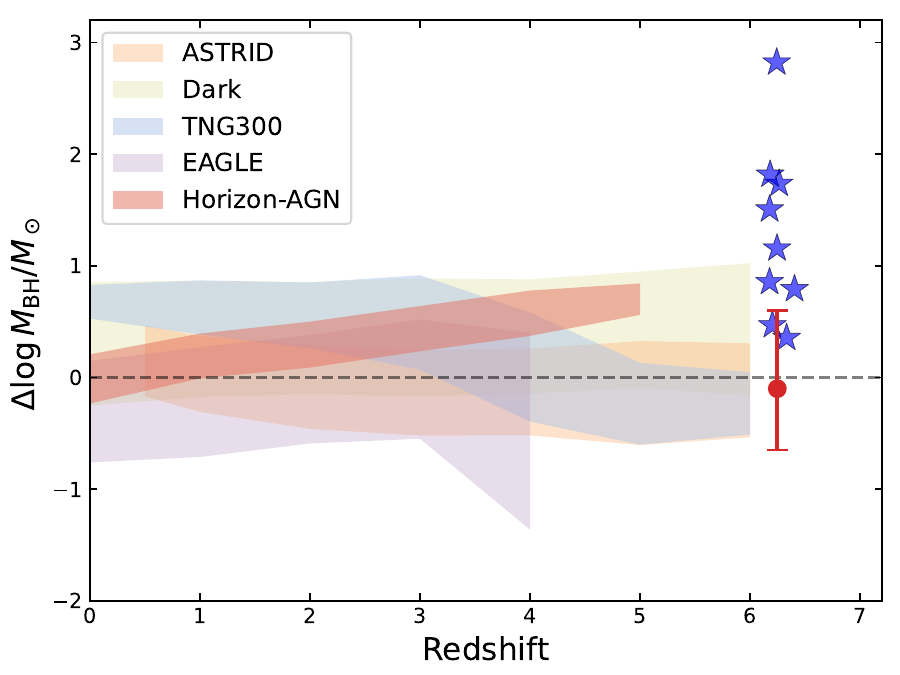}
\caption{Best-fit value and $1\sigma$ confidence interval of the intrinsic mass ratio (red data point), relative to the local \citet{Greene2020} relation, for the ensemble of 9 SHELLQs-JWST quasars. The observed mass ratios for the individual quasars are shown as blue stars. The results from five cosmological hydrodynamic simulations and semi-analytic models at $\log \MS/\Msun\sim10.5$ are indicated \citep{Habouzit2021,Dattathri2024} between 16th and 84th percentiles.}
\label{fig:sims}
\end{figure}

As introduced earlier, the high redshift universe offers an early environment to disentangle the physical processes responsible for the growth of SMBHs and their host galaxies. The observational data at $z>6$ may be used to discriminate between the importance of various input physics implemented in cosmological simulations including the seeding of black holes, their subsequent growth, and feedback processes for SNe and AGN. With this in mind, we compare our results in Figure~\ref{fig:sims} with four cosmological simulations of galaxy and black hole growth, namely ASTRID \citep{Ni2022}, TNG300, \citep{Pillepich2018} Horizon-AGN \citep{Volonteri2016} and EAGLE \citep{Schaye2015}, and the Dark Sage semi-analytic model \citep{Stevens2016,Stevens2018}. Since these simulations do not have a sufficiently large volume to produce BHs with masses comparable to those powering the most massive high-redshift quasars, our comparison is necessarily based on an extrapolation from lower-mass BHs.

The mass ratios vary by 1 to 2 dex across the different simulations and semi-analytic models, highlighting the lack of consensus on whether BHs are, on average, more or less massive at high redshift compared to low redshift. In other words, the offsets indicate whether BH growth proceeds more rapidly or more slowly than the assembly of their host galaxies at early cosmic times. Comparison with observations of high redshift quasars can therefore diagnose a combination of physical processes such as BH initial mass and growth, feedback processes from SNe and AGN.

Simulations such as ASTRID show a \MBH--\MS relation that remains relatively unchanged in both normalization and shape with redshift. BHs and their host galaxies evolve in a similar manner across different redshifts. In contrast, models like Horizon-AGN exhibit smaller mass offsets at lower redshift, which can be attributed to galaxies growing more rapidly than their central BHs at later epochs.  For TNG300, the growth of BHs in low-mass galaxies is more strongly suppressed at high redshift compared to low redshift, due to the modeling of supernova (SN) feedback. This results in an increase in the mass ratios as redshift decreases.

In Figure~\ref{fig:sims}, we compare the BH–to–stellar mass ratios of our individual quasars (observed; blue stars) and the intrinsic mass ratio (red circle and error bar) to the simulations at $\log \MS/\Msun\sim10.5$ (including inactive black holes) and up to $z\sim6$ \citep{Dattathri2024, Habouzit2021}. The 16th--84th percentile ranges indicate the intrinsic scatter in the mass offset for each model. As shown, the variations in the mass ratio between the models are generally similar to the observed $1\sigma$ confidence interval on our best inference on the intrinsic mass ratio at $z>6$. It's worth highlighting once again how closely our measurement matches that of the local relation of \citet{Greene2020}. As evident, a larger observed sample at $z\gtrsim6$ is needed along the inclusion of lower redshift samples \citep[e.g.,][]{Ding2020ApJ...896..159D,Ding2022a,Tanaka2024} to place substantive constraints on current simulations and models.

\section{Summary}

A determination of the evolution in the intrinsic mass relation between SMBHs and their host galaxies is needed to understand the coupling and mutual influence on their growth. While much effort has been achieved at lower redshifts ($z\lesssim2$), the universe at $z>6$ represents early phases of SMBH growth, likely more sensitive to the initial conditions (i.e., seeding and feedback mechanisms).

Towards this goal, we use a sample of nine quasars at $z>6$ from the SHELLQs survey, observed with JWST, in conjunction with an established forward modeling approach to account for selection biases and measurement uncertainties to assess the intrinsic mass relation between SMBHs and their host galaxies. Black hole masses are based on NIRSpec observations of the broad Balmer lines (Onoue et al. in preparation) while the stellar masses of their host galaxies are measured from the 2D decomposition of NIRCam images \citep{Ding2023,Ding2025}. We assume that the observed location of these nine quasars in the \MBH--\MS plane is representative of the larger sample of SHELLQs quasars from \citet{Matsuoka2018ApJ}. Our forward modeling of the relation between SMBHs and their hosts depends on knowledge of the stellar mass function of galaxies and the Eddington rate distribution. Output parameters are the normalization ($\alpha$) and scatter ($\sigma$) of the intrinsic mass relation along with the AGN fraction ($p_\mathrm{active}$). The slope ($\beta$) of the mass relation is fixed to the local value of 1.61 \citep{Greene2020} due to the small sample over a limited range in mass. 

Based on our best understanding of the data and selection effects, we find that the intrinsic mass relation between SMBHs and their stellar mass is unlikely to be elevated much above the local mass relation. Rather, observational selection effects and measurement uncertainties result in the observed offsets seen in the \MBH--\MS plane. Based on a fixed slope in the mass relation, the intrinsic relation at $z>6$ is consistent with the local relation of \citet{Greene2020} which may indicate a scenario where SMBHs and the total stellar mass of their host galaxies grow in tandem.  

Interestingly, the dispersion ($0.80_{-0.28}^{+0.23}$\,dex) in the mass relation at $z>6$ is higher than the local value ($\sim$0.3--0.4\,dex) of their likely more massive descendants. Evolution in the dispersion has not been seen in studies at lower redshifts, particularly based on a large SDSS quasar sample at $z<1$ with Subaru's Hyper Suprime-Cam ($\sigma=0.25_{-0.04}^{+0.03}$ dex; \citealt{Li2021mass}) and AGN at $1\lesssim z <\lesssim2$ with JWST ($\sigma=0.38_{-0.09}^{+0.12}$ dex; \citealt{Tanaka2024}). This new result may support a non-causal connection between galaxies and their supermassive black holes, i.e., a natural outcome of cosmic averaging due to mergers \citep{Jahnke2011}. However, there is uncertainty on the level of scatter (along with the normalization) due a degeneracy with the active fraction, which requires better constraints. Furthermore, we recognize that the dispersion in mass relations is dependent on galaxy type thus these conclusions depend on the choice of the local samples employed. A more focused analysis based on these results and others will be presented in Tanaka et al. (in preparation).

With a non-evolving mass relation, our analysis supports the prediction of \citet{Li2025b} for the existence of a large population of lower mass black holes to be discovered with deeper spectroscopy or different observing strategies. \citet{Li2025a} find lower mass black holes, consistent with the local mass relation, in a search of the JWST archive of high-$z$ massive galaxies with NIRSpec spectroscopy. \cite{Geris2025} further identified a population of low-mass black holes located within the scatter of the local relation.
Also, Ren et al. (submitted) identify candidates below the local relation based on NIRSpec IFU observations of typical star-forming galaxies from the ALPINE-CRISTAL survey. These are likely just the start in expanding the parameter space in black hole and galaxy mass for larger samples of SMBHs with JWST, which will allow the slope to be measured directly thus placing tighter constraints on the normalization and intrinsic scatter of the mass relation.



\section*{Acknowledgements}

J.S. led the writing of the paper and co-led the SHELLQs-JWST program. J. L. is responsible for the modeling analysis,  preparation of figures, and contributions to the text. Measurements of stellar mass based on image decomposition was carried out by X.D. while black hole masses were determined by M.O. The remainder of the team participated in the interpretation of the results and review of the manuscript. We thank the anonymous referee for their insightful review and Jenny Greene for input on the local dispersion in the mass relation.

This work is based on observations made with the NASA/ESA/CSA James Webb Space Telescope. The data were obtained from the Mikulski Archive for Space Telescopes at the Space Telescope Science Institute, which is operated by the Association of Universities for Research in Astronomy, Inc., under NASA contract NAS 5-03127 for JWST. These observations are associated with programs GO \#1967, and GO \#3859. Support for these programs was provided by NASA through a grant from the Space Telescope Science Institute, which is operated by the Association of Universities for Research in Astronomy, Inc.,
under NASA contract NAS 5-03127. This work was supported by World Premier International Research Center Initiative (WPI), MEXT, Japan. 
This work used computing resources at Kavli IPMU.
JS is supported by JSPS KAKENHI (JP22H01262).
MO is supported by the Japan Society for the Promotion of Science (JSPS) KAKENHI Grant No. 24K22894.
YM was supported by the Japan Society for the Promotion of Science (JSPS) KAKENHI Grant No. 21H04494.
MV gratefully acknowledges financial support from the Independent Research Fund Denmark via grant numbers DFF 8021-00130 and  3103-00146 and from the Carlsberg Foundation via grant CF23-0417.
SEIB is supported by the Deutsche Forschungsgemeinschaft (DFG) under Emmy Noether grant number BO 5771/1-1.
KI acknowledges support from the National Natural Science Foundation of China (12073003, 11721303, 11991052).
KI acknowledges support under the grant PID2022-136827NB-C44 provided by MCIN/AEI/10.13039/501100011033 / FEDER, UE.
AL acknowledges support from PRIN MUR 2022 - Project “2022935STW"
JTS is supported by the Deutsche Forschungsgemeinschaft (DFG, German Research Foundation) - Project number 518006966.
FW acknowledges support from NSF award AST-2513040.
MH acknowledges support from the FNS under the SNSF Starting Grant 218032.
BT acknowledges support from the European Research Council (ERC) under the European Union's Horizon 2020 research and innovation program (grant agreement number 950533) and from  the Excellence Cluster ORIGINS, which is funded by the Deutsche Forschungsgemeinschaft (DFG, German Research Foundation) under Germany's Excellence Strategy - EXC 2094 - 390783311.

\bibliography{jdsrefs}{}
\bibliographystyle{aasjournal}





\end{document}